\documentclass [12pt]{article}
\usepackage {graphicx}
\usepackage {longtable}

\begin{document}
\begin{center}
{\LARGE Kerr Solution Consistent Motion of Spin Particles in the General Relativity}
\end{center}

\bigskip

\begin{center}
\textbf{M.V. Gorbatenko\footnote{E-mail: {gorbatenko@vniief.ru}} and T.M. Gorbatenko}
\end{center}

\bigskip

\begin{center}
Russian Federal Nuclear Center - All-Russian Research Institute of 
Experimental Physics, Sarov, Nizhni Novgorod region
\end{center}

\bigskip

\begin{center}
\textbf{Abstract}
\end{center}

\bigskip

The paper presents equations determining the particle spin evolution in the 
post-Newtonian approximation in the problem of motion of two mass and spin 
possessing particles. The equations are derived with the 
Einstein-Infeld-Hoffmann method from the condition of metric tensor 
symmetry. The consideration uses the condition of coordinate harmonicity and 
the metric coincidence nearby the particles with expansions of the Kerr 
solution written in the harmonic coordinates. For gyroscopes on satellites 
Gravity Probe B, for example, the equations yield a deviation of the axis of 
revolution, which is, first, two times as small as that obtained by 
J.L.Anderson in paper gr-qc/0511093 and, second, the deviation is of 
opposite sense. From the equations it follows that the total angular 
momentum of the spin particle system, generally speaking, is not conserved, 
beginning even with the post-Newtonian approximation. The results are 
briefly discussed.

\newpage 

\bigskip

\subsection*{1. Introduction}

\bigskip

A system of two particles possessing masses and intrinsic angular momentums 
is considered. The description of motion of these particles within the 
general relativity requires two types of dynamic equations: the ones 
determining the motion of center of masses of particles and spin evolution 
for each of them. 

The equations of the first type can be derived either with Fock method or 
Einstein-Infeld-Hoffmann (EIH) method. As a rule, no disagreements in regard 
to these equations arise. 

The situation regarding the spin evolution describing equations is 
absolutely different. There are several versions of these equations for 
self-consistent motion of two spin particles (e.g., [1]-[4]) and plenty of 
versions of equations describing the spin dynamics of the trial spin 
particle (e.g., [5]-[17]). All the versions differ from each other, and it 
is far from being evident what of these versions is proper.

We have solved the problem of self-consistent motion of two spin particles 
in the post-Newtonian approximation (PN\footnote{ By the PN approximation is 
meant an approximation, where corrections to the Newtonian approximation 
begin to appear irrespective of the order of smallness of the corrections.} 
) with the EIH method. The distinctive features of our consideration are the 
following.

1. All the operations provided for by the EIH method are performed using de 
Donder condition for metric

\begin{equation}
\label{eq1}
\left( {\sqrt { - g} \;g^{\lambda \sigma} } \right)_{,\sigma}  = 0,
\end{equation}

\noindent
which coincides with the condition of coordinate harmonicity.

2. In the approximation expressions of the EIH method the unknown 
coefficients involve expansions of the Kerr solution written in harmonic 
coordinates in [18], [19]. The Kerr solution expansion coefficients are 
taken from ref. [20]. 

3. Dynamic equations for the spins result from the conditions of the metric 
symmetry in the approximations under consideration. 

The dynamic equations for particle spins that have been derived by us and 
the ones for trial spin particle which follow from them differ from all 
others in that they are consistent with the starting general relativity 
postulate of metric tensor symmetry. Ensuring the metric symmetry, we 
encounter an alternative: either the metric symmetry or the law of 
conservation of angular momentum. As a result of the choice made we arrive 
at nonconservation of the vector of total angular momentum of the particle 
system even in the PN approximation. This issue is interesting by itself, 
and we discuss it briefly. Besides, estimations for gyroscopes, the 
experiments with which were conducted on satellite Gravity Probe B (GP-B), 
are presented and discussed.

We guess that the reader is familiar with the EIH method, so we restrict 
ourselves to minimum elucidations in this regard. Also, cumbersome 
treatments on EIH technique are not included in this paper. All these 
treatments and calculation details will be presented in a separate paper.

\subsection*{2. Notation}

\bigskip

This paper solves the equations of general relativity with zero 
energy-mo\-men\-tum tensor. Like in refs. [21], [22], write the equations using 
$\eta _{\alpha \beta}  $, the metric tensor of Minkowski space.

\begin{equation}
\label{eq2}
R_{\alpha \beta}  - \frac{{1}}{{2}}\eta _{\alpha \beta}  \left( {\eta ^{\mu 
\nu} R_{\mu \nu} }  \right) = 0.
\end{equation}

Here $R_{\alpha \beta}  $ is Ricci tensor specified in the standard manner. 
In the background space the Cartesian coordinates will be used, therefore

\begin{equation}
\label{eq3}
\eta _{\alpha \beta}  = \eta ^{\alpha \beta}  = diag\left[ { - 1,1,1,1} 
\right].
\end{equation}

Two types of quantities are used: $h_{\alpha \beta}  $, $\gamma _{\alpha 
\beta}  $. These quantities are given by:

\begin{equation}
\label{eq4}
g_{\alpha \beta}  = \eta _{\alpha \beta}  + h_{\alpha \beta}  ,
\quad
\gamma _{\alpha \beta}  = h_{\alpha \beta}  - \frac{{1}}{{2}}\eta _{\alpha 
\beta}  \left( {\eta ^{\mu \nu} h_{\mu \nu} }  \right).
\end{equation}

The raising and lowering of index in $h_{\alpha \beta}  $ and $\gamma 
_{\alpha \beta}  $ is made using tensors (\ref{eq3}). Thus, by $h^{\alpha \beta} $ 
is meant $h^{\alpha \beta}  \equiv \eta ^{\alpha \mu} \eta ^{\beta \nu 
}h_{\mu \nu}  $. 

A system of two particles is considered. In the background space we choose 
Cartesian coordinates $\left( {ct,x_{k}}  \right)$ in an arbitrary way. 
Denote the particle radius-vector coordinates by $\xi _{k} ;\eta _{k} $, 
respectively. Denote the length of the vector with coordinates $R_{k} = \eta 
_{k} - \xi _{k} $ by $R$. Denote the derivative with respect to $ct$ either 
by point above the quantity (for example, $\dot {\xi} _{k} $) or as 
subscript ``0'' separated with comma (for example, $\xi _{k,0} \equiv \dot 
{\xi} _{k} $). 

The masses of the first and second particles are denoted by $\tilde 
{M},\tilde {m}$, respectively. Quantities

\[
M = \frac{{G\tilde {M}}}{{c^{2}}};\quad m = \frac{{G\tilde {m}}}{{c^{2}}}
\]

\noindent
are equal to halves of Schwarzschild radii of the particles. Here $G$ is the 
gravitational constant, $c$ is light speed. In what follows quantities $M,m$ 
will be referred to as masses, although they have the length dimension. 
Write the axial vectors of the intrinsic angular momentums of the particles 
in form

\[
\tilde {M}cS_{k} ;\quad \tilde {m}cs_{k} .
\]

The $S_{k} ;s_{k} $ have the length dimension; we will call them the reduced 
vectors of intrinsic angular momentums or simply the spins.

Equations (\ref{eq2}) in the $\lambda ^{k}$-order are denoted as $\left[ {00;\lambda 
^{2}} \right]$, $\left[ {0k;\lambda ^{3}} \right]$, $\left[ {mn;\lambda 
^{4}} \right]$, etc. Hereinafter the coordinate conditions are denoted as 
$\left[ {c.c.;0;\lambda ^{3}} \right]$, $\left[ {c.c.;k;\lambda ^{4}} 
\right]$, etc.

\subsection*{3. Smallness parameters}

\bigskip

In the EIH method the principal smallness parameter is

\begin{equation}
\label{eq5}
\lambda = {{v} \mathord{\left/ {\vphantom {{v} {c}}} \right. 
\kern-\nulldelimiterspace} {c}}
\end{equation}

\noindent
where $v$ is the characteristic relative particle velocity. Alongside the 
parameter $\lambda $, the system under discussion has four more 
dimensionless and a priori independent parameters:

\begin{equation}
\label{eq6}
{{u \equiv M} \mathord{\left/ {\vphantom {{u \equiv M} {r}}} \right. 
\kern-\nulldelimiterspace} {r}};
\quad
{{v \equiv S} \mathord{\left/ {\vphantom {{v \equiv S} {r}}} \right. 
\kern-\nulldelimiterspace} {r}};
\quad
{{m} \mathord{\left/ {\vphantom {{m} {r}}} \right. 
\kern-\nulldelimiterspace} {r}};
\quad
{{s} \mathord{\left/ {\vphantom {{s} {r}}} \right. 
\kern-\nulldelimiterspace} {r}}.
\end{equation}

With large enough values of the radial variable $r$ these parameters become 
small. However, the relative relationships among them are arbitrary in 
principle as they are independent of $r$ and depend on the particle 
characteristics. In this paper the consideration will be under the following 
assumptions:

\begin{equation}
\label{eq7}
{{M}\over{r}}\sim\lambda^2\hskip12mm
{{m}\over{r}}\sim\lambda^2,
\end{equation}

\begin{equation}
\label{eq8}
{{S}\over{r}}\sim\lambda;\hskip12mm
{{s}\over{r}}\sim\lambda.
\end{equation}

We will not dwell on elucidation of the physical meaning of these 
assumptions; these elucidations are presented in many papers and are of 
standard nature. Note only that in many real situations assumptions (\ref{eq7}), (\ref{eq8}) 
are satisfied not in full measure. Nevertheless, the dynamic equations 
derived under assumptions (\ref{eq7}), (\ref{eq8}) can be used as a source of information 
about the motion of spin particles in these cases as well.

\subsection*{4. Arrangement of orders of smallness}

\bigskip

Certain assumptions as to the arrangement of the orders of smallness of some 
terms appearing in functions $\gamma _{00} ,\;\gamma _{0k} ,\;\gamma _{mn} $ 
should be made. The lowest-order approximation term construction is 
determined uniquely by the assortment of parameters (\ref{eq6}). The orders of 
smallness of the other terms are taken so that the EIH procedure be 
self-consistent. Our consideration will use the following arrangements of 
the orders of smallness:

\begin{equation}
\label{eq9}
\left. {\begin{array}{l}
 {\gamma _{00} = \mathop {\gamma} \limits_{2} {}_{00} + \mathop {\gamma 
}\limits_{4} {}_{00} + \mathop {\gamma} \limits_{5} {}_{00} + ...,} \\ 
 {\gamma _{0k} = \mathop {\gamma} \limits_{3} {}_{0k} + \mathop {\gamma 
}\limits_{5} {}_{0k} + \mathop {\gamma} \limits_{6} {}_{0k} + ...,} \\ 
 {\gamma _{mn} = \mathop {\gamma} \limits_{4} {}_{mn} + \mathop {\gamma 
}\limits_{5} {}_{mn} + \mathop {\gamma} \limits_{6} {}_{mn} + ...\;.} \\ 
 \end{array}}  \right\}
\end{equation}

Typically, in the EIH method, the expansions of functions $\gamma _{00} 
,\gamma _{0k} ,\gamma _{mn} $ are composed of terms of the same parity in 
$\lambda $. In our case the results of [20] are taken into consideration, 
from which it follows that the expansions of functions $\gamma _{00} ,\gamma 
_{0k} ,\gamma _{mn} $ should contain terms of all orders of smallness, 
beginning with $\mathop {\gamma} \limits_{4} {}_{00} ,\mathop {\gamma 
}\limits_{5} {}_{0k} ,\mathop {\gamma} \limits_{4} {}_{mn} $, respectively. 

Recall some ``techniques'' used in the EIH method. 

$ \bullet $ In the EIH method, in the differentiation of any function in 
$x^{0}$ its order of smallness increases by one as against the one that the 
function had before the differentiation. In the differentiation with respect 
to spatial coordinates $x^{k}$ the function order of smallness is preserved.

$ \bullet $ All the expressions must be symmetric about the replacement of 
the first particle by the second and the second by the first, that is about 
simultaneous replacements:

\begin{equation}
\label{eq10}
\left. {\begin{array}{l}
 {M \to m;\quad m \to M;\quad S \to s;\quad s \to S;} \\ 
 {\xi _{k} \to \eta _{k} ;\quad \eta _{k} \to \xi _{k} ;\quad R_{k} \to - 
R_{k} .} \\ 
 \end{array}}  \right\}
\end{equation}

$ \bullet $ With vanishing parameters $m,s$ the expressions derived with the 
EIH method should be the same as the associated Kerr solution expansions 
written for the particle with parameters $M,S$. 

$ \bullet $ If inequalities (\ref{eq7}), (\ref{eq8}) hold for the first particle and the 
parameters of the second particle satisfy inequalities

\begin{equation}
\label{eq11}
{{m}\over{r}}\sim\lambda^4;\hskip12mm
{{s}\over{r}}\sim\lambda^3,
\end{equation}

\noindent
then in the PN approximation the second particle can be considered as a 
trial particle. In this case the presence of the second particle in the 
system impacts in no way on the dynamic equations for the first particle; as 
for the second particle, it moves in the field of gravity generated by the 
first particle like in the external field. In so doing the second particle 
is a trial spin particle. 

Input equations (\ref{eq2}) are written in each order of approximation in the form 
of one equation $00$, three equations $0k$, and six equations $mn$. The 
consistency of arrangement of orders of smallness (\ref{eq9}) with equations (\ref{eq2}) 
shows up in the fact that written-out equation chain $\left[ {00;\lambda 
^{2k}} \right]$, $\left[ {0k;\lambda ^{2k + 1}} \right]$, $\left[ 
{mn;\lambda ^{2k + 2}} \right]$ allows us to sequentially determine $\mathop 
{\gamma} \limits_{2k} {}_{00} ,\;\mathop {\gamma} \limits_{2k + 1} {}_{0k} 
,\;\mathop {\gamma} \limits_{2k + 2} {}_{mn} $. It is therewith assumed that 
coordinate conditions consistent with this procedure are used.

\subsection*{5. Explicit form of the Kerr solution expansion}

\bigskip

Let us present the explicit form of the principal terms of the expansions of 
$\gamma _{\alpha \beta}  $ in smallness parameters $u = {{m} \mathord{\left/ 
{\vphantom {{m} {r}}} \right. \kern-\nulldelimiterspace} {r}},\;{{v = a} 
\mathord{\left/ {\vphantom {{v = a} {r}}} \right. \kern-\nulldelimiterspace} 
{r}}$, where $a \equiv \sqrt {\left( {s_{k} s_{k}}  \right)} $. The 
expansions are borrowed from ref. [20]. In the expressions for $\gamma _{00} 
$, $\gamma _{0k} $, $\gamma _{mn} $ we restrict ourselves to writing-out 
only those terms that contain multipliers given by $u$, $u^{2}$. $vu$, 
$vu^{2}$, $v^{2}u$, $v^{3}u$. 

In the Kerr solution, the tensor of intrinsic angular momentum reduced to 
unit mass, $s_{mn} $, has as few as one nonzero component $s_{12} $, whose 
value coincides with $a$. Thus,

\begin{equation}
\label{eq12}
s_{12} = a;\quad \quad s_{23} = s_{31} = 0;\quad \quad s_{1} = s_{2} = 
0;\quad \quad s_{3} = a.
\end{equation}

Axial momentum vector $s_{k} $ is related to tensor $s_{mn} $ by the 
standard relation

\begin{equation}
\label{eq13}
s_{mn} = \varepsilon _{mnc} s_{c} ;\quad s_{c} = \frac{{1}}{{2}}\varepsilon 
_{cab} s_{ab} ;\quad s_{3} = a.
\end{equation}

From (\ref{eq12}), (\ref{eq13}) it follows that the following relations take place:

\begin{equation}
\label{eq14}
1 = \frac{{\left( {s_{c} s_{c}}  \right)}}{{a^{2}}}.
\end{equation}

\begin{equation}
\label{eq15}
cos^{2}\theta = \frac{{\left( {s_{a} x_{a}}  \right)\left( {s_{b} x_{b}}  
\right)}}{{a^{2}r^{2}}}.
\end{equation}

The expression for $\gamma _{00} $ in Cartesian coordinates is given by:

\begin{equation}
\label{eq16}
\begin{array}{l}
 \gamma _{00} = 4u + u^{2} - 6\frac{{s_{a} s_{b}} }{{a^{2}}}\left( 
{\frac{{x_{a} x_{b}} }{{r^{2}}} - \frac{{1}}{{3}}\delta _{ab}}  \right) 
\cdot v^{2} \cdot u + ... \\ 
 = 4\frac{{m}}{{r}} + \frac{{m^{2}}}{{r^{2}}} - 6\frac{{m}}{{r^{3}}}s_{a} 
s_{b} \left( {\frac{{x_{a} x_{b}} }{{r^{2}}} - \frac{{1}}{{3}}\delta _{ab}}  
\right) + ...\;. \\ 
 \end{array}
\end{equation}

The expressions for $\gamma _{01} ,\;\gamma _{02} ,\;\gamma _{03} $ in 
Cartesian coordinates are given by:

\begin{equation}
\label{eq17}
\begin{array}{l}
 \gamma _{0k} = 2\frac{{m\left( {s_{ka} x_{a}}  \right)}}{{r^{3}}} \\ 
 - 2\frac{{m^{2}\left( {s_{ka} x_{a}}  \right)}}{{r^{4}}} - 5\frac{{m\left( 
{s_{kc} x_{c}}  \right)\left( {s_{a} x_{a}}  \right)\left( {s_{b} x_{b}}  
\right)}}{{r^{7}}} + \frac{{m\left( {s_{ka} x_{a}}  \right)}}{{r^{5}}}\left( 
{s_{a} s_{a}}  \right) + ...\;. \\ 
 \end{array}
\end{equation}

The expressions for $\gamma _{11} ,\;\gamma _{22} ,\;\gamma _{33} ,\gamma 
_{12} ,\;\gamma _{13} ,\;\gamma _{23} $ in Cartesian coordinates are given 
by:

\begin{equation}
\label{eq18}
\gamma _{mn} = \frac{{m^{2}x_{m} x_{n}} }{{r^{4}}} - 
2\frac{{m^{2}}}{{r^{2}}}\delta _{mn} + \frac{{m^{2}\left[ {\left( {s_{mc} 
x_{c}}  \right)x_{n} + \left( {s_{nc} x_{c}}  \right)x_{m}}  
\right]}}{{r^{5}}} + ....
\end{equation}

With vanishing $s_{mn} $ expressions (\ref{eq16}) - (\ref{eq18}) transfer to Schwarzschild 
solution expansions.

\subsection*{6. Construction of the Einstein equation solution corresponding to two Kerr 
particles}

\bigskip

\subsubsection*{6.1. An approach to using the EIH method in the problem of two Kerr 
particles}

\bigskip

The approach differs basically in nothing from the one used for two 
particles having nonzero mass, but possessing no intrinsic angular momentums 
(Schwarzschild particles). The problem specificity consists in the 
following: 

\underline {First}, in all orders of smallness functions $\gamma _{00} 
,\;\gamma _{0k} ,\;\gamma _{mn} $ split into two parts,

\begin{equation}
\label{eq19}
\left. {\begin{array}{l}
 {\gamma _{00} = \hat {\gamma} _{00} + \bar {\gamma} _{00}}  \\ 
 {\gamma _{0k} = \hat {\gamma} _{0k} + \bar {\gamma} _{0k}}  \\ 
 {\gamma _{mn} = \hat {\gamma} {}_{mn} + \bar {\gamma} _{mn}}  \\ 
 \end{array}}  \right\} \quad .
\end{equation}

The first parts (with circumflex accent) are solutions to the associated 
problem for two Schwarzschild particles. The second parts (with bar) are 
additions that owe their origin to the particle intrinsic momentums. 
Splitting (\ref{eq19}) can be performed always without loss of generality. 

\underline {Second}, each of the parts of function (\ref{eq19}) is expanded in 
smallness parameters so that they coincide with the exact Kerr solution 
expansions in the two smallness parameters used in the EIH procedure. 
(Recall that in the problem of two Schwarzschild particles one smallness 
parameter was used.) The following arrangement of orders of smallness will 
be used according to exact solution expansions (\ref{eq9}):

\begin{equation}
\label{eq20}
\left. {\begin{array}{l}
 {\gamma _{00} = \left[ {\mathop {\hat {\gamma} }\limits_{2} {}_{00} + 
\mathop {\hat {\gamma} }\limits_{4} {}_{00} + \mathop {\hat {\gamma 
}}\limits_{6} {}_{00} + ...} \right] + \left[ {\mathop {\bar {\gamma 
}}\limits_{4} {}_{00} + \mathop {\bar {\gamma} }\limits_{6} {}_{00} + ...} 
\right]} \\ 
 {\gamma _{0k} = \left[ {\mathop {\hat {\gamma} }\limits_{3} {}_{0k} + 
\mathop {\hat {\gamma} }\limits_{5} {}_{0k} + ...} \right] + \left[ {\mathop 
{\bar {\gamma} }\limits_{3} {}_{0k} + \mathop {\bar {\gamma} }\limits_{5} 
{}_{0k} + ...} \right]} \\ 
 {\gamma _{mn} = \left[ {\mathop {\hat {\gamma} }\limits_{4} {}_{mn} + 
\mathop {\hat {\gamma} }\limits_{6} {}_{mn} + ...} \right] + \left[ {\mathop 
{\bar {\gamma} }\limits_{4} {}_{mn} + \mathop {\bar {\gamma} }\limits_{5} 
{}_{mn} + \mathop {\bar {\gamma} }\limits_{6} {}_{mn} + ...} \right]} \\ 
 \end{array}}  \right\} \quad .
\end{equation}

In the expansion of $\gamma _{mn} $ for the exact solution the odd order 
appears in the $\lambda ^{5}$ approximation.

\underline {Third}, the exact solutions are written with the same coordinate 
conditions as in the EIH procedure. In our consideration these will be the 
harmonic coordinate conditions. 

\underline {Fourth}, functions (\ref{eq20}) are substituted both into the dynamic 
equations and coordinate conditions and the dynamic equations and coordinate 
conditions, which the second parts of the functions (with circumflex accent) 
have to satisfy, are determined. 

\underline {Fifth}, the resultant equations and conditions are solved. In so 
doing the conditions of solvability of the dynamic equations and coordinate 
conditions are satisfied as well.

\subsubsection*{6.2. Determination of $\mathop {\gamma} \limits_{5} {}_{0k} $}

\bigskip

The EIH procedures can be applied essentially in their standard form until 
$\mathop {\gamma} \limits_{5} {}_{0k} $ has to be determined. We will not 
write out the associated expressions as they are cumbersome. These 
expressions along with the details of their derivation will be presented in 
a preprint to be published. 

In this section we only present the explicit expression for $\mathop {\gamma 
}\limits_{5} {}_{0k} $. The function $\mathop {\gamma} \limits_{5} {}_{0k} $ 
is determined from equation $\left[ {0k;\lambda ^{5}} \right]$ and 
coordinate condition $\left[ {c.c.;0k;\lambda ^{5}} \right]$. The 
calculation of surface integrals from the coordinate condition leads to the 
following solvability conditions:

\begin{equation}
\label{eq21}
\mathop {M}\limits_{4} = \frac{{1}}{{2}}M\left( {\dot {\xi} _{l} \dot {\xi 
}_{l}}  \right) - \frac{{1}}{{2}}\frac{{mM}}{{R}};\quad \mathop 
{m}\limits_{4} = \frac{{1}}{{2}}m\left( {\dot {\eta} _{l} \dot {\eta} _{l}}  
\right) - \frac{{1}}{{2}}\frac{{mM}}{{R}}.
\end{equation}

As a result we arrive at:

\begin{equation}
\label{eq22}
\begin{array}{l}
 \mathop {\hat {\gamma} }\limits_{5} {}_{0k} \quad \mathop { = 
}\limits_{\kappa \ge 0} \quad + 8\frac{{mM\dot {\eta} _{k}} }{{Rr_{1}} } - 
4\frac{{mM\dot {\xi} _{k}} }{{Rr_{1}} } - 4\frac{{M\left( {\dot {\xi} _{l} 
\dot {\xi} _{l}}  \right)\dot {\xi} _{k}} }{{r_{1}} } + 2\frac{{M\left( 
{X_{l} \dot {\xi} _{l}}  \right)^{2}\dot {\xi} _{k}} }{{r_{1}^{3}} } \\ 
 - \frac{{M^{2}\left( {X_{l} \dot {\xi} _{l}}  \right)X_{k}} }{{r_{1}^{4}} } 
+ \frac{{M^{2}\dot {\xi} _{k}} }{{r_{1}^{2}} } + 2\frac{{mM\left( {X_{l} 
R_{l}}  \right)\dot {\xi} _{k}} }{{R^{3}r_{1}} } - 2\frac{{mM\left( {X_{l} 
\dot {\xi} _{l}}  \right)R_{k}} }{{R^{3}r_{1}} } \\ 
 - 6\frac{{mM\left( {R_{l} \dot {\eta} _{l}}  \right)X_{k}} }{{R^{3}r_{1}} } 
+ 8\frac{{mM\left( {X_{l} \dot {\eta} _{l}}  \right)R_{k}} }{{R^{3}r_{1}} } 
+ 8\frac{{mM\left( {R_{l} \dot {\xi} _{l}}  \right)X_{k}} }{{R^{3}r_{1}} }. 
\\ 
 \end{array}
\end{equation}

\begin{equation}
\label{eq23}
\begin{array}{l}
 \mathop {\bar {\gamma} }\limits_{5} {}_{0k} \quad \mathop { = 
}\limits_{\kappa \ge 0} \quad 2\frac{{M\left( {\mathop {S}\limits_{3} {}_{ka} 
X_{a}}  \right)}}{{r_{1}^{3}} } - 5\frac{{M\left( {S_{kc} X_{c}}  
\right)\left( {S_{a} X_{a}}  \right)\left( {S_{b} X_{b}}  
\right)}}{{r_{1}^{7}} } + \frac{{M\left( {S_{ka} X_{a}}  
\right)}}{{r_{1}^{5}} }\left( {S_{a} S_{a}}  \right) \\ 
 + 6\frac{{M\left( {S_{l} X_{l}}  \right)\left( {X_{l} \dot {\xi} _{l}}  
\right)S_{k}} }{{r_{1}^{5}} } - 2\frac{{M\left( {S_{l} \dot {\xi} _{l}}  
\right)S_{k}} }{{r_{1}^{3}} } + 2\frac{{M\left( {X_{l} \dot {\xi} _{l}}  
\right)\left( {S_{kl} \dot {\xi} _{l}}  \right)}}{{r_{1}^{3}} } \\ 
 + 4\frac{{mM\left( {s_{kl} R_{l}}  \right)}}{{R^{3}r_{1}} } + 
\frac{{mM\left( {S_{kl} R_{l}}  \right)}}{{R^{3}r_{1}} } - 3\frac{{M\left( 
{X_{l} \dot {\xi} _{l}}  \right)^{2}\left( {S_{kl} X_{l}}  
\right)}}{{r_{1}^{5}} } \\ 
 - 2\frac{{M^{2}\left( {S_{kl} X_{l}}  \right)}}{{r_{1}^{4}} } - 
\frac{{mM\left( {X_{l} R_{l}}  \right)\left( {S_{kl} X_{l}}  
\right)}}{{R^{3}r_{1}^{3}} } + 4\frac{{mM\left( {X_{a} S_{ab} R_{b}}  
\right)X_{k}} }{{R^{3}r_{1}^{3}} } \\ 
 + 4\frac{{mM\left( {s_{kl} X_{l}}  \right)}}{{R^{3}r_{1}} } + 
12\frac{{mM\left( {X_{a} s_{ab} R_{b}}  \right)R_{k}} }{{R^{5}r_{1}} } + 
6\frac{{mM\left( {X_{l} R_{l}}  \right)\left( {X_{a} S_{ab} R_{b}}  
\right)X_{k}} }{{R^{5}r_{1}^{3}} } \\ 
 + 6\frac{{mM\left( {X_{a} S_{ab} R_{b}}  \right)R_{k}} }{{R^{5}r_{1}} } + 
4\frac{{mM\left( {S_{kl} X_{l}}  \right)}}{{R^{3}r_{1}} } - 6\frac{{mM\left( 
{X_{l} R_{l}}  \right)\left( {S_{kl} R_{l}}  \right)}}{{R^{5}r_{1}} }. \\ 
 \end{array}
\end{equation}

\subsection*{7. A method for derivation of the dynamic equations in the PN approximation}

\bigskip

The above expression for $\mathop {\gamma} \limits_{5} {}_{0k} $ is needed to 
derive the dynamic equations determining the motion of centers of masses of 
particles as well as equations determining time derivatives of spins, i.e. 
$\mathop {\dot {S}}\limits_{3} {}_{mn} $ and $\mathop {\dot {s}}\limits_{3} 
{}_{mn} $. 

The expression for $\mathop {\gamma} \limits_{5} {}_{0k} $ involves not 
$\mathop {\dot {S}}\limits_{3} {}_{mn} $, $\mathop {\dot {s}}\limits_{3} 
{}_{mn} $, but $\mathop {S}\limits_{3} {}_{mn} $, $\mathop {s}\limits_{3} 
{}_{mn} $. The equations for $\mathop {\dot {S}}\limits_{3} {}_{mn} $ and 
$\mathop {\dot {s}}\limits_{3} {}_{mn} $ appear after $\mathop {\gamma 
}\limits_{5} {}_{0k} $ has been substituted into the coordinate condition 
relating $\mathop {\gamma} \limits_{5} {}_{0k,0} $ and $\mathop {\gamma 
}\limits_{6} {}_{ks,s} $, and this condition is used to derive $\mathop 
{\gamma} \limits_{6} {}_{mn} $. But we need not determine the complete 
expression for $\mathop {\gamma} \limits_{6} {}_{mn} $, it will suffice to 
only determine the anti-symmetric part of $\mathop {\gamma} \limits_{6} 
{}_{mn} $ and require that it vanish. It is this condition that will give the 
dynamic equations for the spin evolution. 

The equations of translational motion of particles are derived as before 
through integration of equation $\left[ {mn;\lambda ^{6}} \right]$. A new 
point is that, first, a different decomposition of the terms of this 
equation to the curl combination and the part denoted by$\sum\limits_{i = 
1}^{18} {\alpha _{i}}  $ (a different amount of $\alpha _{i} $ and a 
different explicit form of them) takes place. Second, in the integration of 
$\sum\limits_{i = 1}^{18} {\alpha _{i}}  $ it should be kept in mind that 
the contributions are made not only by Schwarzschild parts $\hat {\gamma 
}_{\mu \nu}  $, but also by Kerr parts $\bar {\gamma} _{\mu \nu}  $. The 
strategy of derivation of the equations of translational motion of particles 
consists in determination of $\hat {\gamma} _{\mu \nu}  $, $\bar {\gamma 
}_{\mu \nu}  $, which make a contribution to the integral of $\left[ 
{mn;\lambda ^{6}} \right]$, and in calculation of the integral. Note that 
the contribution of $\mathop {\gamma} \limits_{6} {}_{mn} $ to the integral 
is zero, so for this purpose $\mathop {\gamma} \limits_{6} {}_{mn} $ need not 
be determined in the explicit form.

The resultant dynamic equations determining the motion of center of masses 
of particles coincide with those presented in many papers, e.g., [1]. 
Therefore, we do not write them out. Note only that the laws of conservation 
of energy and particle momentum vector follow from these equations.

\bigskip

\subsection*{8. The dynamic equations for spins}

\bigskip

The resultant equations determining the spin dynamics that have been derived 
by the above method are given by:

\begin{equation}
\label{eq24}
\mathop {\dot {S}}\limits_{3} {}_{c} = 9\frac{{m}}{{R^{3}}}\left( {R_{l} \dot 
{\xi} _{l}}  \right)S_{c} + 2\frac{{m}}{{R^{3}}}\left( {R_{l} \dot {\eta 
}_{l}}  \right)S_{c} - 2\frac{{m}}{{R^{3}}}\left( {S_{l} \dot {\xi} _{l}}  
\right)R_{c} .
\end{equation}

\begin{equation}
\label{eq25}
\mathop {\dot {s}}\limits_{3} {}_{k} = - 9\frac{{M}}{{R^{3}}}\left( {R_{l} 
\dot {\eta} _{l}}  \right)s_{k} - 2\frac{{M}}{{R^{3}}}\left( {R_{l} \dot 
{\xi} _{l}}  \right)s_{k} + 2\frac{{M}}{{R^{3}}}\left( {s_{l} \dot {\eta 
}_{l}}  \right)R_{k} .
\end{equation}

\subsection*{9. Time variation in total momentum}

\bigskip

The substitution of the resultant dynamic equations for spins into the 
equations for time derivative of total momentum of the particle system leads 
to the following expression for $\dot {{\rm M}}_{c} $:

\begin{equation}
\label{eq26}
\begin{array}{l}
 \dot {{\rm M}}_{c} = + \frac{{mM}}{{R^{3}}}\left\{ { - 2\left( {S_{l} \dot 
{\xi} _{l}}  \right)R_{c} + 9\left( {R_{l} \dot {\xi} _{l}}  \right)S_{c} + 
2\left( {R_{l} \dot {\eta} _{l}}  \right)S_{c}}  \right\} \\ 
 - \frac{{mM}}{{R^{3}}}\left\{ { - 2\left( {s_{l} \dot {\eta} _{l}}  
\right)R_{c} + 9\left( {R_{l} \dot {\eta} _{l}}  \right)s_{c} + 2\left( 
{R_{l} \dot {\xi} _{l}}  \right)s_{c}}  \right\} \\ 
 + \varepsilon _{cmn} \left\{ { - 6\frac{{mM\left( {s_{ml} R_{l}}  
\right)\left( {R_{l} \dot {R}_{l}}  \right)}}{{R^{5}}}R_{n} - 
4\frac{{mM\left( {s_{ml} \dot {\xi} _{l}}  \right)}}{{R^{3}}}R_{n} + 
3\frac{{mM\left( {s_{ml} \dot {\eta} _{l}}  \right)}}{{R^{3}}}R_{n}}  
\right\} \\ 
 + \varepsilon _{cmn} \left\{ { - 6\frac{{mM\left( {S_{ml} R_{l}}  
\right)\left( {R_{l} \dot {R}_{l}}  \right)}}{{R^{5}}}R_{n} - 
3\frac{{mM\left( {S_{ml} \dot {\xi} _{l}}  \right)}}{{R^{3}}}R_{n} + 
4\frac{{mM\left( {S_{ml} \dot {\eta} _{l}}  \right)}}{{R^{3}}}R_{n}}  
\right\} \\ 
 - 3\frac{{mM}}{{R^{5}}}\left( {\left( {S_{l} + s_{l}}  \right)R_{l}}  
\right)\left( {S_{cl} + s_{cl}}  \right)R_{l} \\ 
 \end{array}
\end{equation}

The analysis of the structure of the right-hand side of the equation for 
$\dot {{\rm M}}_{c} $ suggests that no re-determination of the total angular 
momentum of the particle system can get the law of conservation of momentum 
to be obeyed in the PN approximation. 

For a spin particle system located in plane-asymptotic background space it 
may be possible to construct a conserved quantity similar to the total 
angular momentum. As it follows from our consideration, however, the 
quantity will not reduce definitely to the sum of orbital moments of the 
particles and their intrinsic angular momentums. 

The obtained result of nonconservation of the total angular momentum of the 
particle system in the PN approximation appears unusual, however, this 
result seems not to contradict the first principles of general relativity. 
Most probably, similar effects will appear with respect to energy and 
momentum as well at some approximation step.

\subsection*{10. Numerical estimations for the gyroscopes on satellite GPB}

\bigskip

For the numerical estimations we will use the input data for gyroscopes that 
are presented in ref. [3]. According to that paper, the gyroscopes were 
spheres of radius $r_{g} = 1.9cm$ and mass $m = 75g$. The initial angular 
velocity was $\omega = 27\;000\;rad/s$. The satellite was launched onto a 
nearly ideal polar orbit of radius $R = 7027\;km$. The orbital velocity of 
the gyroscopes is $v = 7.5 \cdot 10^{5}\;cm/s$. Ratio $v/c$ is ${{v} 
\mathord{\left/ {\vphantom {{v} {c}}} \right. \kern-\nulldelimiterspace} 
{c}} = 2.5 \cdot 10^{ - 5}$. The ratio of half the Schwarzschild radius of 
the Earth to the gyroscope orbit radius is

\begin{equation}
\label{eq27}
\frac{{M}}{{R}} = 6.5 \cdot 10^{ - 10}.
\end{equation}

The gyroscope axes were in the satellite orbit plane.

The equation used in ref. [3] to estimate the gyroscope axis of revolution 
deviation is derived in that paper itself using Landau-Lifshitz 
energy-momentum pseudo-tensor [23]. It is given by:

\begin{equation}
\label{eq28}
\dot {s}_{k} = \frac{{1}}{{5}}\frac{{M}}{{R^{3}}}\left\{ { - \left( {s_{l} 
\dot {R}_{l}}  \right)R_{k} - 19\left( {s_{l} R_{l}}  \right)\dot {R}_{k} + 
16s_{k} \left( {R_{l} \dot {R}_{l}}  \right)} \right\}.
\end{equation}

Superpose the satellite orbit plane to plane $\left( {x,y} \right)$ and 
write the change in radius-vector in form

\begin{equation}
\label{eq29}
R_{1} = Rcos\left( {\Omega t} \right);\quad R_{2} = Rsin\left( {\Omega t} 
\right);\quad \Omega = \frac{{2\pi} }{{T}}.
\end{equation}

From these conditions it follows that

\begin{equation}
\label{eq30}
\dot {\eta} _{1} = - R\Omega sin\left( {\Omega t} \right);\quad \dot {\eta 
}_{2} = R\Omega cos\left( {\Omega t} \right).
\end{equation}

At the initial time the gyroscope axis of revolution orientation coincides 
with axis $x$,

\begin{equation}
\label{eq31}
s_{1\left( {0} \right)} = s_{0} ;\quad s_{21\left( {0} \right)} = 0.
\end{equation}

From Anderson's equation it follows that the contribution to the time 
derivatives from the spin is made by the two first terms in the right-hand 
side of (\ref{eq28}). Upon averaging of these terms over the orbit for the time of 
one revolution we obtain that

\begin{equation}
\label{eq32}
\frac{{\dot {s}_{1}} }{{s_{0}} }T = 0;\quad \frac{{\dot {s}_{2}} }{{s_{0} 
}}T = \frac{{M}}{{R}}.
\end{equation}

Thus, in one revolution around the Earth the gyroscope axes gets displaced 
by angle $\Delta \theta $ of

\begin{equation}
\label{eq33}
\Delta \theta = {{2M} \mathord{\left/ {\vphantom {{2M} {R}}} \right. 
\kern-\nulldelimiterspace} {R}} = 1.27 \cdot 10^{ - 9}.
\end{equation}

In its functioning the satellite GPB had to perform about 5000 revolutions, 
hence, the gyroscope deviation angles of the order of $6 \cdot 10^{ - 
6}\;radian$ had to be measured in the experiment.

From equations (\ref{eq24}), (\ref{eq25}) for spins that were derived with the method of 
this paper it follows that, first, the deviation must be 2 times as small 
and, second, the gyroscope axes deviate to the side opposite to the one 
following from equations (\ref{eq28}).

\subsection*{11. Conclusions}

As far as we know, the procedure of expansion of the Kerr solution written 
in the harmonic coordinates had been performed by nobody before our paper 
[20], hence, had not been used for normalization of expressions derived with 
the EIH method. This entitles us to suggest that the particle spin motion 
results presented elsewhere should be validated through comparison with the 
Kerr solution expansions in Section 5. 

Another test for correctness of the equations for spins that are presented 
in the literature is the absence of any anti-symmetric part in the 
expression for $\mathop {\gamma} \limits_{6} {}_{mn} $, which can appear from 
coordinate condition $\left[ {c.c.;\lambda ^{6}} \right]$. 

Now as for the amount of the gyroscope axis of revolution deviation on 
satellite GP-B. From the consideration presented in Section 10 it follows 
that our equations (\ref{eq24}), (\ref{eq25}) give $\Delta \theta = - M/R$ for the deviation 
(per revolution around the Earth), while equations (\ref{eq28}) give $\Delta \theta 
= + 2M/R$. We guess that the general relativity predicts the effect that 
agrees with our equations (\ref{eq24}), (\ref{eq25}). It is extremely interesting to learn 
the results of the forthcoming gyroscope experiment on satellite Gravity 
Probe B! 

From the obtained results it follows that the case in point should be not 
only the effects of the deviation from motion along the geodesic, precession 
and axis of revolution orientation variation, but also the change in lengths 
of intrinsic momentums, i.e. values of angular velocity of rotation. 
Gyroscope rotation acceleration/deceleration on different segments of the 
world line could be noted in the laboratory without orbiting as well. To do 
this sufficient statistics should be compiled regarding the amount of 
revolutions made by the gyroscope about its axis for one and the same time 
on segments, where the spin projection onto the radius-vector is positive 
and negative. The comparison of the revolution amount on these two segments 
can serve a test for the general relativity validation.

\bigskip

The authors are thankful to L.S. Mkhitaryan for discussion of the results of 
this paper.

\bigskip

\subsection*{References}

\bigskip

[1] V.A.Brumberg. \textit{Relativistic Celestial Mechanics} [in Russian]. 
Nauka Publishers, Moscow (1972).

[2] A.P. Ryabushko. \textit{The motion of bodies in general relativity} [in 
Russian]. Vysheishaya Shkola Puiblishers. Minsk (1979).

[3] J.L.Anderson. \textit{Approximate Equations of Motion for Compact 
Spinning Bodies in General relativity.} ArXiv: gr-qc/0511093.

[4] M.V.Gorbatenko. \textit{Theor. and Math. Physics}. Vol. \textbf{101}, 
1245 (1994).

[5] V.A. Fock. J. Phys. U.S.S.R. \textbf{1}. 81 (1939).

[6] A. Papapetrou. Proc. Phys. Soc. \textbf{64}. 57 (1951).

[7] A. Papapetrou. Proc. Roy. Soc. Lond. \textbf{A209}. 248 (1951).

[8] E. Corinaldesi and A. Papapetrou. Proc. Roy. Soc. Lond. \textbf{A209}. 
259 (1951). 

[9] F. Pirani. Acta Phys. Polon. \textbf{15}. 389 (1956). 

[10] L.I. Schiff. Phys. Rev. Lett. \textbf{4}. 215 (1960). 

[11] L.I. Schiff. Proc. Nat. Acad. Sci. \textbf{46}. 871 (1960).

[12] W.G. Dixon. Proc. Roy. Soc. Lond. \textbf{A314}. 499 (1970). 

[13] S.Weinberg. \textit{Gravitation and Cosmology: Principles and 
Applications of the General Theory Relativity}. New York. Wiley (1972).

[14] Yu.S. Vladimirov. \textit{Frames in gravitational theory} [in Russian]. 
Energoizdat Publishers, Moscow (1982).

[15] D.V.Gal'tsov, V.I.Petukhov and A.N.Aliev. \textit{Phys. Letters.} 
\textbf{105A}, No. 7, 346 (1984).

[16] I.D. Novikov, V.P. Frolov. \textit{Physics of black holes} [in 
Russian]\textit{.} Nauka Publishers, Moscow (1986).

[17] I.B. Khriplovich. \textit{General Relativity Theory} [in 
Russian]\textit{.} Moscow, Institute of Computer Studies (2002).

[18] A.A.Vlasov, A.A.Logunov. \textit{Theor. and Math. Physics.} Vol. 
\textbf{70}, 171 (1987). 

[19] A.A. Logunov, M.A. Mestvirishvili. \textit{Relativistic Gravitational 
Theory} [in Russian]\textit{.} Nauka Publishers, Moscow (1989).

[20] M.V.Gorbatenko, T.M.Gorbatenko. \textit{Theor. and Math. Physics.} Vol. 
\textbf{140}, 1028 (2004).

[21] A.Einstein, L.Infeld, and B.Hoffmann. \textit{Gravitational Equations 
and Problems of Motion.} Ann. Math. \textbf{39}, 65-100 (1938).

[22] A.Einstein, L.Infeld. \textit{On the Motion of Particles in General 
Relativity Theory}. Can. J. Math. \textbf{1}, 209-241 (1949).

[23] L.D. Landau, E.M. Lifshitz. \textit{Field Theory} [in 
Russian]\textit{.} Nauka Publishers, Moscow (1988).

\end{document}